# Optimization of ARQ Protocols in Interference Networks with QoS Constraints

Marco Levorato*, Daniel O'Neill*, Andrea Goldsmith* and Urbashi Mitra†

* Dept. of Electrical Engineering, Stanford University, Stanford, CA 94305 USA
† Dept. of Electrical Engineering, University of Southern California, Los Angeles, USA.
e-mail: levorato@stanford.edu, ubli@usc.edu, andreag@stanford.edu.

*Abstract*—We study optimal transmission strategies in interfering wireless networks, under Quality of Service constraints. A buffered, dynamic network with multiple sources is considered, and sources use a retransmission strategy in order to improve packet delivery probability. The optimization problem is formulated as a Markov Decision Process, where constraints and objective functions are ratios of time-averaged cost functions. The optimal strategy is found as the solution of a Linear Fractional Program, where the optimization variables are the steady-state probability of state-action pairs. Numerical results illustrate the dependence of optimal transmission/interference strategies on the constraints imposed on the network.

## I. INTRODUCTION

Retransmission-based error control techniques have been widely employed to improve reliability to communications against the impairments of the wireless channel [1]–[3]. In time-varying channels, the transmission of multiple copies of a packet can provide diversity and improve the Quality of Service (QoS) of the link. Implementations of retransmission-based error control techniques range from pure Automatic Retransmission reQuest (ARQ), where packets are sent uncoded over the channel, to hybrid ARQ, which introduces packet encoding and memory of previous transmissions [4]–[6].

ARQ techniques have been mostly studied in single link scenarios [1]–[6]. This paper studies ARQ in interference networks, where multiple sources may access the same time-frequency resource. Mutual interference couples the behavior and effectiveness of link level ARQ protocols. This, in turn, couples the stochastic evolution of the content of each link's buffer. For example, two links simultaneously transmitting can adversely effect the packet error probability of each link and thus through the ARQ protocol, the contents of each link's buffer.

The coupling between interference and ARQ process has been studied in cognitive networks, where the ARQ protocol of the primary sources is fixed [7], [8]. In this paper, we instead center the discussion on the optimization of multiple and mutually inter-dependent retransmission processes with QoS constraints.

We consider a network of multiple sources with packet arrival, buffering and memoryless retransmission-based error control. The network is modeled as a collection of inter-dependent stochastic processes. A constrained infinite-horizon Markov Decision Process (MDP) is formulated in order to optimize the transmission/interference strategy of the sources. Performance metrics such as packet delivery probability, average throughput, total packet delay and unit of energy spent per unit of throughput are the objective/constraint functions of the optimization problem.

The MDP is solved through a linear fractional program, where the optimization variables are the steady-state probability of state-action pairs. Optimizing the ratio of time averaged cost functions yields optimal retransmission strategies. In fact, the formalization as a linear fractional program enables an easy incorporation in the optimization problem of individual packet performance and relevant tradeoffs (*e.g.*, energy per unit of throughput, average delay over failure probability), which are obtained as ratio of time-averages of cost functions defined on the state-action space. To the best of our knowledge, this is the first formalization of an MDP problem incorporating these objective/constraints. Interestingly, the proposed framework finds connections with optimization frameworks used to minimize the cost per unit of time in controlled semi-Markov processes [9], [10].

The observation of the optimal transmission/interference strategies enables the understanding of objective/constraints related behaviors, which may serve as guidelines for practical protocols. Numerical results are provided for a network with two sources with the goal of minimizing the aggregate average energy per unit of throughput with constraints on individual source's throughput, individual packet total delay and failure probability.

The remainder of the paper is organized as follows. In Section II the considered network is described. Section III defines the stochastic model of the network, the performance metrics and the optimization problem. In Section IV the linear fractional program used to solve the constrained infinite-horizon MDP addressed in this paper is described in detail. Section V provides a renewal interpretation of some performance metrics. Section VI investigates the optimal strategy for an instantiation of the network.

## II. NETWORK DESCRIPTION

A single-hop network of $S$ sources is considered. Each source $s=1,\ldots,S$ stores packets to be delivered to its intended destination in a finite First-In First-Out buffer of size $B$ packets.

Sources adopt a memoryless ARQ retransmission-strategy in order to improve packet delivery probability. Therefore, prior transmissions of the wanted packet are discarded at the receiver. More refined retransmission protocols providing combination of packets referring to the same information content, such as type-II hybrid ARQ, can be incorporated in the model at the price of a larger state space of the stochastic model.

We fix a maximum time interval for packet service. The transmission/interference strategy, then, defines packet retrans-

$$\mathcal{P}(Y_k{=}y_k|X_k{=}x_k,...,X_0{=}x_0,Y_{t-1}{=}y_{t-1},...,Y_0{=}y_0,U_k{=}u_k,...,U_0{=}u_0){=}\mathcal{P}(Y_k{=}y_k|X_k{=}x_k,U_k{=}u_k){=}P(y_k|x_k,u_k). \quad (2)$$

mission within this interval. The service interval of a packet is defined as the time elapsed since it became the oldest in the queue and the time it is removed form the buffer. Time slotted operations are assumed, where the duration of the transmission of a packet plus its associated ARQ feedback fits with the duration of one time slot. The maximum service time, then, is fixed to $F$ slots, which corresponds to the maximum number of transmissions of a packet. A packet is removed from the buffer either if successfully delivered to the intended destination or has been in service for $F$ slots.

Packet arrival in the buffer of each individual source is modeled through the variable $\alpha_s$, denoting the probability that a new packet arrives in the buffer of source $s$ in a slot. This simple model is used to enable the obtaining of a clear relationship between the transmission/interference strategy and the queue/service time state of the network. More involved packet arrival processes (*e.g.*, Markovian arrivals) can be easily incorporated into the framework.

The sources' transmission/interference strategy is the solution of an offline optimization problem that maximizes a performance metric subject to QoS constraints. In particular, the optimization problem is formalized as a constrained infinite-horizon undiscounted MDP, where the optimal policy controls packet transmission and dropping at each individual source.

The next section describes, in detail, the stochastic model of the network, the selected performance metrics and the MDP. The formulation of the linear fractional program used to solve the optimization problem is provided in Section IV.

III. STOCHASTIC MODEL AND PERFORMANCE METRICS

The network is modeled as collection of random processes and control sequences tracking individual sources' state (queue length and service time) and actions (packet transmission and dropping from the buffer). Interference ties together the stochastic processes of the individual sources. In fact, the success probability of a source's transmission depends on the set of sources which transmit in the time slot. Therefore, other sources' activity determines the probability that a packet is removed from the queue due to successful delivery or experiences continued service because of a failed transmission. Moreover, in the case considered, the optimal policy is a randomized stationary policy (see Section IV), in which the probability that an action is chosen is a function of the overall state of the network.

In order to characterize the performance of the aggregate network and of the individual sources, a set of cost functions mapping the state-action space to a real cost is defined. The performance metrics, are in turn defined as ratios of time-averages of those cost functions. As explained later in this section, this construction enables the formalization of individual packet and individual source performance, as well as relevant tradeoffs, required to accurately track the performance of the retransmission and channel access strategy.

*A. Stochastic Model of the Network*

Consider the homogeneous random processes $\boldsymbol{X} = \{X_0, X_1, X_2, ...\}$ and $\boldsymbol{Y} = \{Y_0, Y_1, Y_2, ...\}$, where $X_k$ and $Y_k$, $k = 0, 1, 2, ...$, take values in the finite state spaces $\mathcal{X}$ and $\mathcal{Y}$, respectively. We also define the control sequence $\boldsymbol{U}=\{U_0, U_1, U_2, ...\}$, where the control variables $U_k$, $k=0,1,2,...$, take values in the finite action set $\mathcal{U}$. Process $\boldsymbol{X}$ models the correlated temporal evolution of the network given the control sequence $\boldsymbol{U}$, whereas process $\boldsymbol{Y}$ represents a sequence of random outcomes of state-action pairs.

The transition probabilities of $\boldsymbol{X}$ are denoted by

$$P(x_{k+1}|x_k, y_k, u_k) = \mathcal{P}(X_{k+1}{=}x_{k+1}|X_k{=}x_k, Y_k{=}y_k, U_k{=}u_k), \quad (1)$$

where $\mathcal{P}(\cdot)$ denotes the probability of an event. The probability that $Y_k$ takes a certain value $y$ does not depend on the past history of $\boldsymbol{X}$ and $\boldsymbol{Y}$, but only on the action variable $u$ and on the current state of process $X_k$ (see Eq. (2)). The probability that $\boldsymbol{X}$ moves from state $x_k$ to $x_{k+1}$ conditioned on action $u_k$ is

$$P(x_{k+1}|x_k, u_k) = \sum_{y_k \in \mathcal{Y}} P(x_{k+1}|x_k, y_k, u_k) P(y_k|x_k, u_k). \quad (3)$$

The process $\boldsymbol{X}$ tracks the state of the sources in terms of queue length and service time. In particular, the state of $\boldsymbol{X}$ at time $t$ is decomposed into $S$ variables $X_k(s)$, $s=1,...S$, with $X_k(s) \in \mathcal{X}(s) = 0 \cup \{1,...,F\} \times \{1,...,B\}$.[1] $X_k(s)=0$ means that source $s$ has an empty buffer, whereas $X_k(s)=\{b_k(s), f_k(s)\}$, with $f_k(s)=1,...,F$ and $b_k(s)=1,...,B$, means that source $s$ has $b_k(s)$ packets in its buffer and the packet currently under service has been served for $f_k(s)$ slots.

The policy $\mu$ controls sources' access and packet dropping from the buffer. Assuming causal control, $\mu$ is a function of the past states of the processes and control variables, *i.e.*, $u_k=\mu(x_0,...,x_{k-1}, y_0,...,y_{k-1}, u_0,...,u_{k-1})$. However, for the optimization problem formalized in the following, there exists an optimal *randomized stationary* policy [11]. The control variable $U_k$ can be split into individual source variables $U_k(s)$, $s=1,...,S$, determining source $s$'s transmission and packet dropping in the time slot $t$.[2] In particular, $U_k(s)=(T_k(s), D_k(s))$, where $T_k(s)=1$ and $T_k(s)=0$ correspond to transmission and idleness in slot $k$, respectively, and $D_k(s)=1$ and $D_k(s)=0$ correspond to packet dropping and permanence in the buffer of the packet currently being served. Note that if $X_k(s)=0$, *i.e.*, source $s$ has an empty buffer, then $T_k(s)$ and $D_k(s)$ are forced to zero. Moreover, $D_k(s)$ is forced to one if $X_k(s)=(b_k(s), F)$, as the packet currently under service is always dropped after $F$ slots.

---

[1] We recall that $B$ is the size of the buffer and $F$ is the maximum service time

[2] Power control, and in general any transmission parameter, can be included in the model by extending the set $\mathcal{U}$.

The random process $Y$ tracks the success/failure of all the sources of the network. In particular, $Y_k$ takes values in $\mathcal{Y}=\{0,1\}^S$. Again, the variable $y \in \mathcal{Y}$ is decomposed into multiple variables $y(s) \in \{0,1\}$. $y(s)=0$ and $y(s)=1$ corresponds to failure and success of source $s$'s transmission, respectively. The success probability of source $s$ in state $x$ given that action $u$ is chosen is denoted by $\rho_s(x,u)=P(y_k(s)=1|x_k,u_k)$. If $T(s)=0$, *i.e.*, source $s$ is idle in slot $k$, then $\rho_s(x,u)=0$.

## B. Performance Metrics and Optimization Problem

Much of prior work on optimization of transmission scheduling focused on performance metrics such as throughput [12]–[14]. Alternatively, packet delay can be constrained using Lyapunov functions [15]. In order to characterize the performance of individual source and individual packet transmission, we propose the construction of specific objective/constraints functions defined as ratios of time-averages of cost functions.

In particular, we define the set of cost functions $z_a : \mathcal{X} \times \mathcal{Y} \times \mathcal{U} \to \mathbb{R}$, $a=1,\ldots,A$, which assign to the triple $(x,y,u)$ a finite cost $z_a(x,y,u)$, for any $x \in X$, $y \in Y$ and $u \in \mathcal{U}$. The time-average of the cost function $z_a$ is defined as

$$\overline{z}_a(\boldsymbol{U}) = \limsup_{n \to +\infty} \frac{1}{n} \sum_{k=1}^{n} E\big[z_a(X_k, Y_k, U_k)\big], \quad (4)$$

$a=1,\ldots,A$, where $E[\cdot]$ is the expectation operator.

The objective and constraint functions are defined as ratios of time-averages of cost functions.

$$R(\boldsymbol{U}) = \overline{z}_{r_n}(\boldsymbol{U})/\overline{z}_{r_d}(\boldsymbol{U}) \quad (5)$$
$$C_q(\boldsymbol{U}) = \beta_q \, \overline{z}_{a_n(q)}(\boldsymbol{U})/\overline{z}_{a_d(q)}(\boldsymbol{U}) + \lambda_q, \quad (6)$$

respectively, where $\lambda_q$ and $\beta_q$ are constants in $\mathbb{R}$, and $r_n$, $r_d$, $a_n(q)$ and $a_d(q)$ are indexes in $1,\ldots,A$.

The optimization problem is the determination of the sequence $\widehat{\boldsymbol{U}}$ that minimizes the objective function $R(\boldsymbol{U})$ over all the control sequences in $\mathcal{U}^\infty$ subject to $M_c$ constraints on the functions $z_q(\boldsymbol{U})$. Formally,

$$\widehat{\boldsymbol{U}} = \arg\inf_{\boldsymbol{U} \in \mathcal{U}^\infty} R(\boldsymbol{U}) \quad (7)$$
$$\text{s.t.} \quad C_q(\boldsymbol{U}) \leq \gamma_q, \quad \text{for} \quad q=1,2,\ldots,M_c.$$

The above optimization problem represents a constrained infinite-horizon Markov Decision Process (MDP).

Assuming $\boldsymbol{X}$ is *unichain*, *i.e.*, the transition matrix for any stationary deterministic policy has a single recurrent class plus a (perhaps empty) set of transient states [16], then there exists an optimal stationary randomized policy solving the above optimization problem [11]. Moreover, the optimal policy has at most $M_c$ *randomizations*, *i.e.*, states in which the policy is non-deterministic [11].

In the following, the performance metrics used to characterize the performance of the network are listed. The average normalized throughput of the source $s$ is the time-average of the cost function

$$z_1(x_k, y_k, u_k) = \begin{cases} \rho_s(x_k, u_k) & \text{if } u_k(s) : t_k(s)=1, \\ 0 & \text{otherwise,} \end{cases} \quad (8)$$

where $x_k(s)=(t_k(s),d_k(s))$. Similarly, the average normalized energy expense of source $s$ is the time average of the cost function

$$z_2(x_k, y_k, u_k) = t_k(s). \quad (9)$$

Note that the aggregate normalized throughput and energy expense can be obtained as sum of the individual source throughput and energy expense. The ratio $\overline{z}_2/\overline{z}_1$ measures the efficiency of source's $s$ transmission/interference strategy in terms of unit of energy spent per unit of delivered traffic.

Individual packet performance metrics such as packet success probability, number of transmissions and total delay can be obtained as ratios of time-averages of apposite cost functions. In the first two metrics, the number of delivered packets or overall transmissions needs to be normalized to the number of effectively served packets, which is function of the policy. The average total delay, *i.e.*, the average time a packet spends in the buffer, is computed as the ratio between the average queue level and the average number of packet arrivals.

The fraction of slots in which source $s$ successfully delivers a packet to the intended destination is $\overline{z}_4$, where

$$z_3(x_k, y_k, u_k) = \begin{cases} 1 & \text{if } y_k(s)=1, \\ 0 & \text{otherwise.} \end{cases} \quad (10)$$

and $\rho_s(x,u)=1$ if $u : t(s)=0$, *i.e.*, source $s$ is idle. The time average $\overline{z}_3$ of the cost function

$$z_4(x_k, y_k, u_k) = \begin{cases} 1 & \text{if } f_k(s)=1, \\ 0 & \text{otherwise.} \end{cases} \quad (11)$$

measures the fraction of time in which source $s$ starts the service of a new packet. The ratio $\overline{z}_3/\overline{z}_4$ corresponds to the average number of packets successfully delivered by source $s$ normalized to the number of served packets, *i.e.*, the success probability of source $s$'s packets. In fact,

$$\frac{\overline{z}_3}{\overline{z}_4} = \frac{\lim_{n \to +\infty} \sup n \sum_{k=1}^{n} E\big[z_3(X_k, Y_k, U_k)\big]}{\lim_{n \to +\infty} \sup n \sum_{k=1}^{n} E\big[z_4(X_k, Y_k, U_k)\big]}. \quad (12)$$

Note that the cost functions $z_4$ and $z_5$ are indicator functions of subsets of the state-action space of the network. In this case, the associated time-averages correspond to a probability measure. In particular, the time average of a cost function sampling the occurrence of a subset of the state-action space is the steady-state probability of the subset. As discussed in detail in Section V, in this case, the ratio of time-averages assumes a particular meaning connected to renewal theory. The average number of transmissions of a packet of source $s$ is $\overline{z}_2/\overline{z}_4$.

According to Little's law [17], the total delay of a packet of source $s$, defined as the average time a packet spends in the buffer of source $s$, can be measured as the ratio $\overline{z}_6/\overline{z}_7$, where

$$z_6(x_k, y_k, u_k) = b_k(s), \quad (13)$$

and

$$z_7(x_k, y_k, u_k) = \begin{cases} \alpha_s & \text{if } x_k(s) : b_k(s) < B, \\ 0 & \text{if } x_k(s) : b_k(s) = B. \end{cases} \quad (14)$$

In fact, $\overline{z}_6/\overline{z}_7$ corresponds to the ratio of the average queue length and the average number of packets arrived in the buffer of source $s$.

## IV. Optimization Framework

The optimal policy is a stationary randomized policy $\mu : \mathcal{X} \times \mathcal{U} \to [0,1]$, where $\mu(x,u)$ indicates the probability that action $u \in \mathcal{U}$ is selected in state $x \in \mathcal{X}$. Given the policy $\mu$, it is possible to define the transition kernel

$$P_\mu(x_{k+1}|x_k) = \sum_{y_k \in \mathcal{Y}, u \in \mathcal{U}} P(x_{k+1}|x_k, y_k, u_k) P(y_k|x_k, u_k) \mu(x_k, u_k), \quad (15)$$

$\forall x_k, x_{k+1} \in \mathcal{X}$, which denotes the probability that $\boldsymbol{X}$ moves from state $x_k$ to state $x_{k+1}$ under policy $\mu$.

Since $\mathcal{X}$ is unichain, then for any $\mu$ as defined above the limiting distribution $\pi_\mu(x) = \lim_{t \to +\infty} P_\mu^t(x|x')$ exists $\forall x', x \in \mathcal{X}$ [16], where $P_\mu^t(x|x')$ is the $t$-step transition probability from state $x'$ to state $x$.[3] We remark that since the limit $\lim_{t \to +\infty} P_\mu^t(x|x')$ converges to $\pi_\mu(x)$, then [18]

$$\pi_\mu(x) = \lim_{m \to +\infty} \frac{1}{m} \sum_{k=0}^{m-1} P_\mu^t(x|x')$$

$$= \lim_{m \to +\infty} \frac{1}{m} \sum_{k=0}^{m-1} E_\mu[1(X_k = x | X_0 = x')], \quad (16)$$

where $E_\mu[\cdot]$ and $1(\cdot)$ are the expectation operator, conditioned on policy $\mu$, and the indicator function. Thus, $\pi_\mu(x)$ is the average fraction of time spent by $\boldsymbol{X}$ is state $x$.[4]

The average cost collected by the network in state $x$ associated with action $u$ is

$$z_a(x,u) = \sum_{y \in \mathcal{Y}} z_a(x,y,u) P(y|x,u). \quad (17)$$

Therefore, the average cost collected by the network in state $x$ under policy $\mu$ is

$$z_a(x,\mu) = \sum_{u \in \mathcal{U}} z_a(x,u) \mu(x,u), \quad a=1,...A. \quad (18)$$

The time averages in Eq. (4) can then be rewritten as the following linear combinations[5]

$$\overline{z}_a(\mu) = \sum_{x \in \mathcal{X}} \pi_\mu(x) z_a(x,\mu), \quad a=1,...A. \quad (19)$$

The optimization problem (7) becomes

$$\widehat{\mu} = \arg\inf_\mu \frac{\sum_{x \in \mathcal{X}} \pi_\mu(x) z_{r_n}(x,\mu)}{\sum_{i \in \mathcal{X}} \pi_\mu(x) z_{r_d}(x,\mu)} \quad (20)$$

$$\text{s.t. } \beta_q \frac{\sum_{i \in \mathcal{X}} \pi_\mu(x) z_{a_n(q)}(x,\mu)}{\sum_{x \in \mathcal{X}} \pi_\mu(x) z_{a_d(q)}(x,\mu)} + \lambda_q \leq \gamma_q, q=1,...,M_c,$$

where $\widehat{\mu}$ denotes the optimal stationary policy. Since $\mathcal{U}$ is finite, and the limiting distribution exists, the above optimization problem can be restated as a Linear Program optimizing over the admissible polyhedron of the steady-state distribution of the state-action pairs [11].

Define the optimization variable $\omega_{x,u}$ as the probability that the process $\boldsymbol{X}$ is in state $x$ *and* action $u$ is chosen. The reward function $R(\mu)$ and the constraint functions $C_q(\mu)$ can then be expressed as

$$R(\mu) = \frac{\sum_{x \in \mathcal{X}} \sum_{u \in \mathcal{U}} z_{r_n}(x,u) \omega_{x,u}}{\sum_{x \in \mathcal{X}} \sum_{u \in \mathcal{U}} z_{r_d}(x,u) z_{i,u}} \quad (21)$$

$$C_q(\mu) = \beta_q \frac{\sum_{x \in \mathcal{X}} \sum_{u \in \mathcal{U}} z_{a_n(q)}(x,u) \omega_{x,u}}{\sum_{x \in \mathcal{X}} \sum_{u \in \mathcal{U}} z_{a_d(q)}(x,u) \omega_{x,u}} + \lambda_q, \quad (22)$$

or equivalently

$$R(\mu) = \boldsymbol{z}_{r_n}^T \boldsymbol{\omega} / \boldsymbol{z}_{r_d}^T \boldsymbol{\omega} \quad (23)$$

$$C_q(\mu) = \beta_q \boldsymbol{z}_{a_n(q)}^T \boldsymbol{\omega} / \boldsymbol{z}_{a_d(q)}^T \boldsymbol{\omega} + \lambda_q, \quad (24)$$

where $T$ denotes the transpose operator, and $\boldsymbol{z}_a = [z_a(x,u)]_{i \in \mathcal{X}, u \in \mathcal{U}}$ and $\boldsymbol{\omega} = [\omega_{x,u}]_{i \in \mathcal{X}, u \in \mathcal{U}}$ are $|\mathcal{X} \times \mathcal{U}|$ column vectors listing the costs and the steady-state probabilities associated with state $x$ and decision $u$, $\forall x, u$.

Note that the constraints can be restated as the following linear combinations of the variables $\boldsymbol{\omega}$

$$(\beta_q \boldsymbol{z}_{a_n(q)} + (\lambda_q - \gamma_q) \boldsymbol{z}_{z_d(q)})^T \boldsymbol{\omega} \leq 0, \ q=1,...,M_c \quad (25)$$

and collected in the matrix form $\boldsymbol{z} \boldsymbol{\omega} \leq 0$ where $\boldsymbol{z}$ is a $M_c \times |\mathcal{X} \times \mathcal{U}|$ matrix. Define, with a slight abuse of notation, $\boldsymbol{P}$ as a $|\mathcal{X}| \times |\mathcal{X} \times \mathcal{U}|$ matrix, such that the element in the column and row corresponding to the pair $(x',u)$ and $x$ is equal to $1 - \mathcal{P}(x|x',u)$ if $x=x'$, and $-\mathcal{P}(x|x',u)$ if $x \neq x'$.

The optimization problem can then be formalized as the following *Linear-fractional Program* [19]

$$\widehat{\boldsymbol{\omega}} = \arg\min_{\boldsymbol{z}} (\boldsymbol{z}_{r_n}^T \boldsymbol{\omega}) / (\boldsymbol{z}_{r_d}^T \boldsymbol{z}) \quad (26)$$

$$\text{s.t. } \boldsymbol{\omega} \boldsymbol{z} \leq \boldsymbol{0}_{M_c,1},$$

$$\begin{bmatrix} \boldsymbol{1}_{1,|\mathcal{X} \times \mathcal{U}|} \\ \boldsymbol{P} \end{bmatrix} \boldsymbol{\omega} = \begin{bmatrix} 1 \\ \boldsymbol{0}_{|\mathcal{X}|,1} \end{bmatrix}$$

$$\omega_{x,u} \geq 0, \ \forall x \in \mathcal{X}, u \in \mathcal{U}$$

where $\boldsymbol{1}_{m,n}$ and $\boldsymbol{0}_{m,n}$ are $m \times n$ matrices whose elements are set to one and zero, respectively.

The equality constraints force $\boldsymbol{z}$ to be an admissible steady-state distribution for the transition probabilities of $\boldsymbol{X}$, and are equivalent to

$$\sum_{x \in \mathcal{X}} \sum_{u \in \mathcal{U}} \omega_{x,u} = 1 \quad (27)$$

$$\sum_{x \in \mathcal{X}} \sum_{u \in \mathcal{U}} P(x|x',u) \omega_{x',u} = \sum_{u \in \mathcal{U}} \omega_{x,u}, \ \forall x. \quad (28)$$

If $\{\boldsymbol{\omega} : \boldsymbol{z} \boldsymbol{\omega} \leq 0, -\boldsymbol{I} \boldsymbol{\omega} \leq 0, \boldsymbol{1}^T \boldsymbol{\omega} = 1, \boldsymbol{P} \boldsymbol{\omega} = 0, \boldsymbol{z}_{r_d}^T \boldsymbol{\omega} > 0\}$ is a feasible set, then the above problem can be easily transformed to the following equivalent linear program via the

---

[3] The $t$-step transition probabilities can be inductively found from $P_\mu(x|x')$ [16].

[4] We underline that, under the hypothesis that all the chain is unichain, the limiting distribution of the chain is independent of the initial state.

[5] In the following notation, the action sequence $\boldsymbol{U}$ is substituted with the function $\mu$.

$$\overline{z}_\phi(\mu) = \lim_{n\to+\infty} \sup \frac{1}{n} E_\mu\left[\sum_{k=1}^n z_a(X_k, Y_k, u_k)\right] = \lim_{n\to+\infty} \sup \frac{1}{n} \sum_{k=1}^n \mathcal{P}_\mu(X_k\in\mathcal{X}_\phi, Y_k\in\mathcal{Y}_\phi, U_k\in\mathcal{U}_\phi|x') \quad (31)$$

$$= \lim_{n\to+\infty} \sup \frac{1}{n} \sum_{k=1}^n \left(\sum_{x\in\mathcal{X}_\phi} \mathcal{P}_\mu^t(X_k=x|X_0=x') \sum_{y\in\mathcal{Y}_\phi, u\in\mathcal{U}_\phi} \mathcal{P}(Y_k=y|X_k=x, U_k=u)\mathcal{P}(U_k=u|X_k=x)\right)$$

$$= \sum_{x\in\mathcal{X}_\phi} \pi_\mu(x) \left(\sum_{y\in\mathcal{Y}_\phi, u\in\mathcal{U}_\phi} \mathcal{P}_\mu(y|x,u)\mu(x,u)\right) \triangleq \pi_\mu(\phi).$$

---

change of variables $\kappa = g\omega$ [19, Ch. 4.2.3]:

$$\{\widehat{\kappa}, \widehat{g}\} = \arg\min_{\kappa, g} z_{r_n}^T \kappa \quad (29)$$
$$\text{s.t.} \quad z\kappa \leq 0_{M_c, 1},$$
$$\begin{bmatrix} 1_{1, |\mathcal{X}\times\mathcal{U}|} \\ z_{r_d}^T \\ P \end{bmatrix} \kappa - \begin{bmatrix} 1 \\ 0 \\ 0_{|\mathcal{X}|, 1} \end{bmatrix} g = \begin{bmatrix} 0 \\ 1 \\ 0_{|\mathcal{X}|, 1} \end{bmatrix}$$
$$g \geq 0, \quad \kappa_{x,u} \geq 0, \quad \forall x\in\mathcal{X}, u\in\mathcal{U}.$$

If $\exists u \in \mathcal{U} : \widehat{\omega}_{x,u} > 0$, then the optimal time-sharing map in state $x$ is $\widehat{\mu}(x, u) = \widehat{\omega}_{x,u}/\sum_{u\in\mathcal{U}} \widehat{\omega}_{x,u}$. If $\sum_{u\in\mathcal{U}} \widehat{\omega}_{x,u}=0$, i.e., $x$ is transient under the optimal policy, then a deterministic action can be chosen at random such that $\widehat{\mu}(x,u)=1$ if $u=u^* \in \mathcal{U}$, and $\widehat{\mu}(x,u)=0$ otherwise.

## V. Renewal Interpretation

In this section, we discuss the case in which the cost functions are used to *sample* the occurrence of a subset of the state-action space. For instance, a cost function indicating the occurrence of the first slot of a service interval of a new packet can be used to measure the number of packets served by a source. The time-average of this cost function corresponds to the fraction of slots in which this specific state occurs, i.e., its steady-state probability.

In this case, the ratio of average functions can be interpreted, as the average number of occurrences of a subset of the state-action space per renewal interval, where the renewal event corresponds to the occurrence of another subset of the state-action space.

Define the *event* $\phi$ as the set $\mathcal{X}_\phi \times \mathcal{Y}_\phi \times \mathcal{U}_\phi$, where $\mathcal{X}_\phi \subseteq \mathcal{X}$, $\mathcal{Y}_\phi \subseteq \mathcal{Y}$ and $\mathcal{U}_\phi \subseteq \mathcal{U}$. We then say that event $\Phi$ occurs at time $t$ if $X_k \in \mathcal{X}_\phi$, $Y_t \in \mathcal{Y}_\phi$ and $u_k \in \mathcal{U}_\phi$. In words, event $\Phi$ occurs at time $t$ if the process $X$ enters the set of states $\mathcal{X}_\phi$, the value of the random process $Y$ belongs to the subset $\mathcal{Y}_\phi$ and an action in $\mathcal{U}_\phi$ is selected. The time-average of the sampling function

$$z_\phi(x_k, y_k, u_k) = \begin{cases} 1 & \text{if } \{x_k, y_k, u_k\} \in \mathcal{X}_\phi \times \mathcal{Y}_\phi \times \mathcal{U}_\phi \\ 0 & \text{otherwise,} \end{cases} \quad (30)$$

measures the average fraction of time in which event $\phi$ occurs.

Note that, in this case, $\overline{z}_\phi(\mu)$ is a probability measure, which we denote by $\pi_\mu(\phi)$ and that corresponds to the probability that $\phi$ occurs in a randomly chosen slot (see Eq. (31)). Moreover, the average time between two consecutive occurrences of $\phi$ is $E_\mu[\tau_\phi(\ell)]=\tau_\phi=1/\pi_\mu(\phi)$ [18], where $\tau_\phi(\ell)$ is the time between the $\ell$-th and the $\ell+1$-th occurrence of event $\phi$.

Consider now two events $\phi$ and $\psi$. Assume $\overline{z}_\psi(\mu)>0$, that is, the number of times the network *hits* event $\psi$ in an infinite sample-path is infinite. The ratio $\overline{z}_\phi(\mu)/\overline{z}_\psi(\mu)=\pi_\mu(\phi)/\pi_\mu(\psi)$ can be used to formulate performance metrics expressed as the average number of occurrences of $\phi$ per occurrence of $\psi$, e.g., average number of transmissions per packet. In other words, $\overline{z}_\phi(\mu)/\overline{z}_\psi(\mu)=\pi_\mu(\phi)/\pi_\mu(\psi)$ is the ratio between the frequencies of the two events.

It can be observed that, due to the characterization of $X$ and $Y$, the occurrence of an event $\psi$ is a *renewal* event [18] for the network, meaning that the future evolution after an occurrence of $\psi$ does not depend on the past history of the processes. As a consequence, the sample path of $X$ and $Y$ can be split into *renewal intervals* [18] defined by the occurrence of $\psi$. The functionals of the states of $X$ and $Y$ computed in any renewal interval have the same distribution.

Define $N_\psi(t)$ as the process counting the occurrences of $\psi$ up to time $t$. The number of occurrences of $\phi$ within the $\ell$-th renewal interval is denoted with the random variable $V_\phi(\ell)$. The cumulative process $W_\phi(\ell)$ is then defined as the sum $W_\phi(\ell)=\sum_{l=1}^\ell V_\phi(l)$. Note that

$$\lim_{\ell\to+\infty} E_\mu[W_\phi(\ell)]/\ell = \lim_{t\to+\infty} E_\mu[N_\phi(t)]/t. \quad (32)$$

The following holds [18]:

$$E_\mu[V_\phi(\ell)]/\tau_\phi = \lim_{\ell\to+\infty} E_\mu[W_\phi(\ell)]/\ell, \quad (33)$$

i.e., the average occurrences of $\phi$ per unit of time in any renewal interval is equal to the average number of occurrences of $\phi$ per unit of time in the whole sample-path of the process.

We observe that, due to the assumption $\overline{z}_\psi(\mu)>0$, $E_\mu[V_\phi(\ell)]$ is finite, and the above limit converges. It follows:

$$E_\mu[V_\phi(\ell)] = \lim_{\ell\to+\infty} \frac{\tau_\psi}{\ell} E_\mu[W_\phi(\ell)] = \frac{\lim_{t\to+\infty} E_\mu[N_\phi(t)]/t}{\lim_{t\to+\infty} E_\mu[N_\psi(t)]/t}$$
$$= \pi_\mu(\phi)/\pi_\mu(\psi), \quad (34)$$

that is, the ratio between the steady-state probabilities of the two events is equal to the average number of occurrences of $\psi$ in a single renewal interval defined by consecutive occurrences of $\psi$.

This observation connects the present work to the framework presented in [9], [10], in which a linear fractional program is used to minimize the average cost per unit of time of a controlled Semi-Markov process. In the framework considered

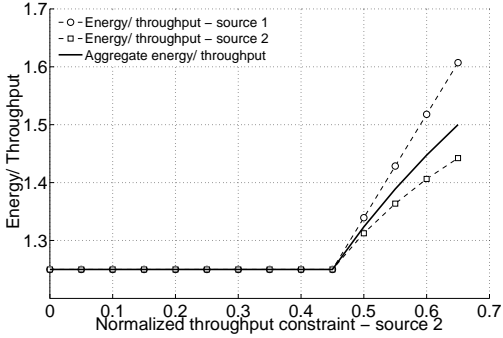
(a) Average aggregate energy/throughput

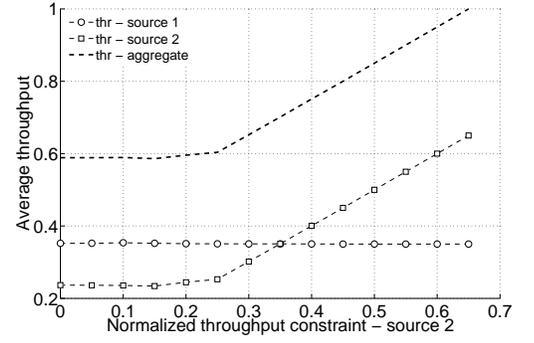
(b) Average throughput

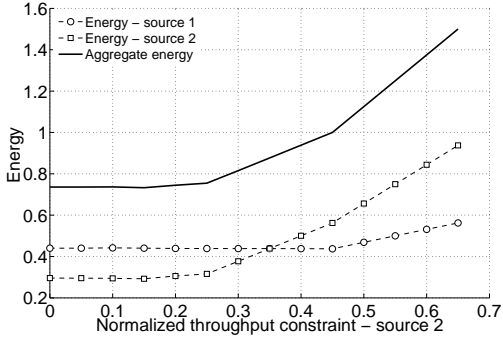
(c) Average energy expense

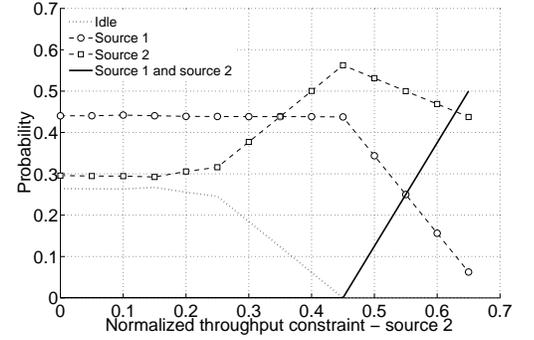
(d) Transmission probability

Fig. 1. Average aggregate energy expense per unit of throughput, average throughput, average energy expense and transmission probability as a function of the constraint on the minimum normalized throughput of source 2.

in [9], [10], to each state of the Markov chain is associated an average time interval. The denominator of the objective function, then, is used to measure the average amount of time the process spends in a state. In the proposed framework, if the cost functions are use to sample the occurrence of a subset of the state-action space, the reference time is the average renewal time, where the renewal intervals are defined by the occurrence of the event associated to the denominator.

## VI. NUMERICAL RESULTS

In this section, we provide numerical results for the framework presented before. In particular, the optimization problem is formalized to minimize the aggregate normalized unit of energy spent per unit of throughput achieved in a two-source network with constraints on the individual source minimum throughput, maximum total delay and maximum packet delivery failure probability (including retransmissions). This setting is motivated by the considerable interest in energy efficient wireless communications of late [20]. As a general observation, stringent QoS constraints force the system to move from time-splitting to simultaneous transmission scheduling. The latter achieves, in the considered setting, a larger throughput and allows a faster packet delivery with respect to time-splitting. On the other hand, simultaneous transmission is less efficient, i.e., requires a larger energy expense per unit of throughput.

A buffer of size $B=1$ is assumed in the first two sets of plots in order to investigate the relation between the transmission/interference strategy and the service time. Note that, in this case, if the state of the individual source $s$ is $x(s) \neq 0$ then $b(s)=1$.

Fig. 1 shows the average aggregate energy expense per unit of throughput, average throughput, average energy expense and transmission probability as a function of the constraint on the minimum normalized throughput of source 2 where the minimum average normalized throughput of source 1 is fixed to $0.35$. The packet arrival probabilities are $\alpha_1=\alpha_2=0.95$. The maximum service time and the buffer size are $F=5$ and $B=1$, respectively. The failure probability of a single source transmitting alone and with interference from the other source are $\rho=0.2$ and $\rho^*=0.4$, respectively. The minimum throughput of source 1 is fixed to $0.35$. The minimum packet delivery probability is $0.8$. The maximum packet total delay is $3.5$ slots.

For the selected parameter setting, the strategy letting only one source to transmit at a given time produces a better energy over throughput balance with respect to the strategy forcing both the sources to transmit. On the other hand, the aggregate throughput achieved with the latter strategy is larger than that associated with the former strategy. Therefore, as long as the throughput requirements are below a certain threshold, the controller allows only a single source to transmit in each slot, and tunes the fraction of slots assigned to each source in order to meet the constraints. It can be observed that, given that only a single source transmits in each slot, the average energy spent per unit of throughput does not depend on how often a source transmits, so that the overall balance remains the same. As

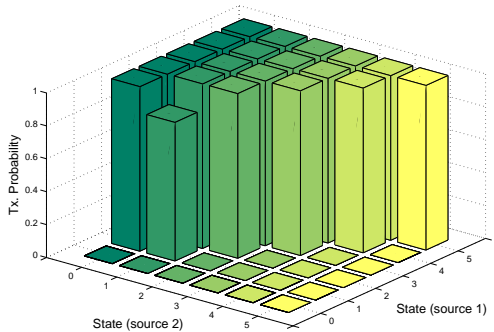
(a) Transmission probability - source 1

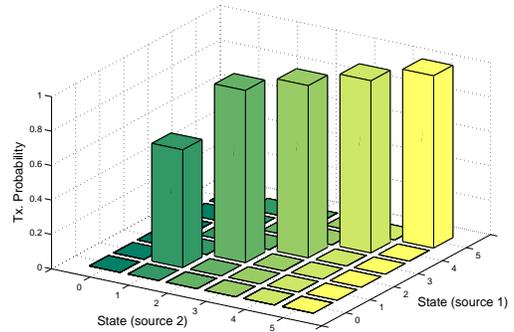
(b) Transmission probability - source 1 and 2

Fig. 2. a) transmission probability of source 1 as a function of the state (the transmission probability of source 2 is symmetric). b) probability of simultaneous transmission of both the sources as a function of the state.

soon as the throughput requirement of source 2 goes above a certain threshold, the controller is forced to let both the sources transmit in a fraction of slots in order to collect a larger throughput, thus worsening the energy/throughput balance. It can be observed that in the region in which the controller schedules simultaneous transmissions, the source with the smallest throughput requirement (source 1 in the figures) is forced to transmit more often than the other source in slots where both the sources transmit, *i.e.*, those slots providing the worst energy/throughput balance (Fig. 1(d)). In fact, source 1 spends more energy to collect a unit of throughput with respect to source 1 (Fig. 1(a)). On the other hand, source 2 is often scheduled in *interference free* slots in order to collect throughput, and achieves a higher energy efficiency. The optimal strategy in this simple configuration suggests that channel access protocols should schedule transmission by sources with relaxed QoS constraints in slots accessed by other sources, while reserving part of the channel resource to sources with stringent QoS constraints. Note that idle time, which is initially scheduled in order to save energy, vanishes for high throughput requirements.

Fig. 2 plots the transmission probability as a function of the state of the network for a similar setting, where the minimum throughput requirement is equal to $0.45$ for both the sources, $\alpha_1=\alpha_2=0.6$ and the total delay constraint is 5slots.

Fig. 2(a) shows the transmission probability of source 1 as a function of the individual state $x(1)$ and $x(2)$,[6]. Interestingly, transmission probability clusters in the state space. In particular, source 1 transmits if $f(1)>f(2)$, *i.e.*, the service time of source 1's packet is larger than that of source 2's packet. This strategy is meant to reduce the probability of packet discarding because of maximum service time expiration. The region of the state space allocated for Source 2's transmission is symmetric to that shown in Fig. 2(a). The probability of simultaneous transmission, shown in Fig. 2(b) is larger than zero on the border-region between the transmission areas of the two sources. Other results, not shown here, indicate that the area of the state space in which simultaneous transmission is scheduled grows as the minimum throughput requirement is increased. This result shows how the state space is split to achieve the largest energy efficiency with stringent throughput requirements.

[6]The queue state $b(s)$ is omitted because either equal to 0 if $x(s)=0$ or 1 if $x(s)>0$.

Fig. 3 and 4 investigates the optimal strategy in a scenario with larger buffer size ($B=3$ and $F=3$). Fig. 3 plots the average aggregate energy expense per unit of throughput, average throughput, average energy expense and transmission probability as a function of the constraint on the maximum total delay of source 2's packets. The maximum total delay of source 1's packets is fixed to 5slots. The minimum throughput requirement is equal to $0.3$ for both the sources. It can be observed that, as the total delay constraint source 2's packets is relaxed, source 1's throughput, transmission probability and energy expense increase, whereas those of source 2 decrease. In fact, a stringent delay constraint forces source 2 to transmit often in order to deliver packets, whereas source 1 is often forced to idleness in order to reduce its impact in terms of interference. Due to the constraint on the delivery probability, which limits packet discarding, delay and throughput are connected. In fact, in order to achieve a smaller delay, sources are forced to transmit, thus increasing the throughput.

Fig. 4(a) and 4(b) plot the energy expense per unit of throughput in the feasible region. In Fig. 4(a), the x and y axis are the throughput constraint of source 1 and 2, respectively. The maximum total delay is 5 slots. In Fig. 4(b), the x and y axis are the throughput and total delay constraint of both source 1 and 2. In general, stringent throughput and total delay constraints require the system to allocate simultaneous transmissions in some regions of the state space. As simultaneous transmission requires a larger energy expense per unit of throughput, the efficiency of the network decreases.

## VII. CONCLUSIONS

We present a general framework to find optimal ARQ strategies. We model the network as a set of three intertwined stochastic processes. The framework extremizes an MDP under constraints, using techniques from Linear Fractional Programming. Different objectives or different constraints will result in different optimal ARQ strategies.

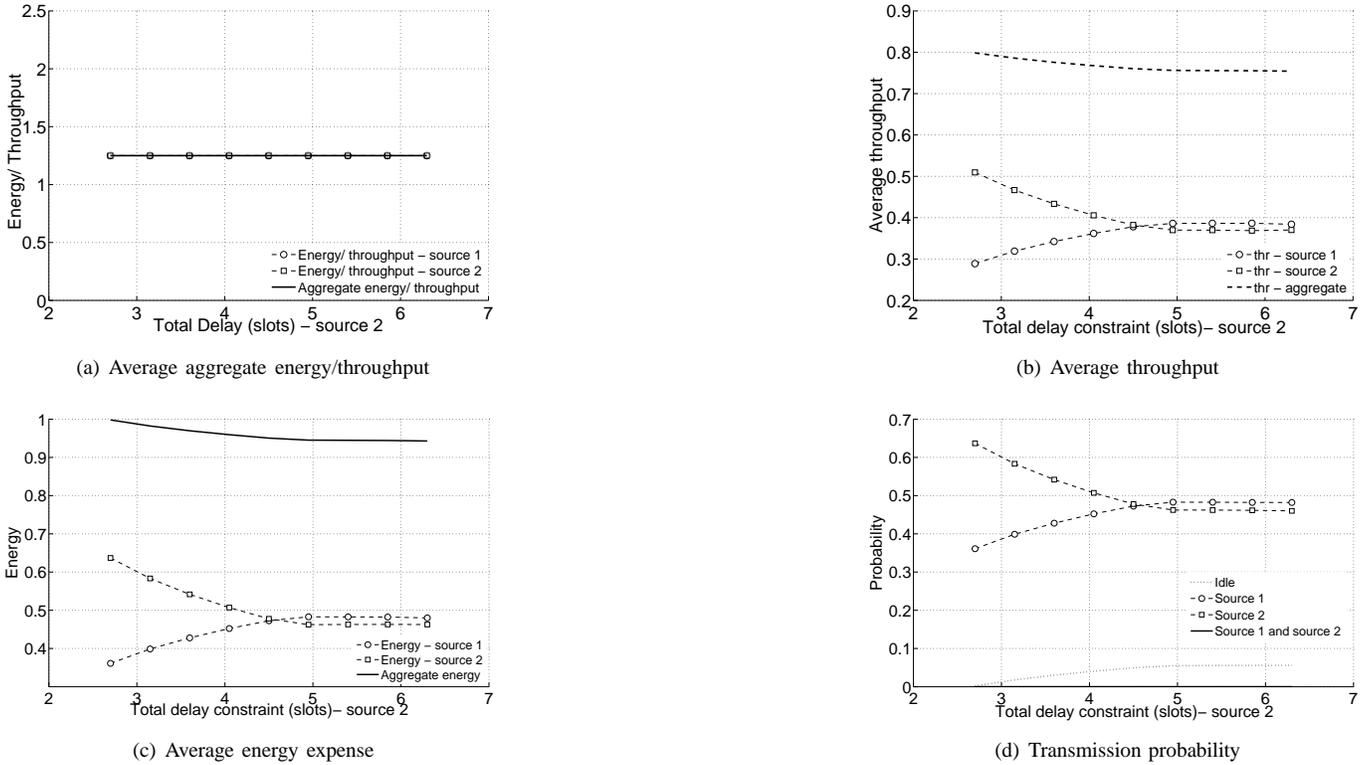

Fig. 3. Average aggregate energy expense per unit of throughput, average throughput, average energy expense and transmission probability as a function of the constraint on the maximum total delay of source 2's packets

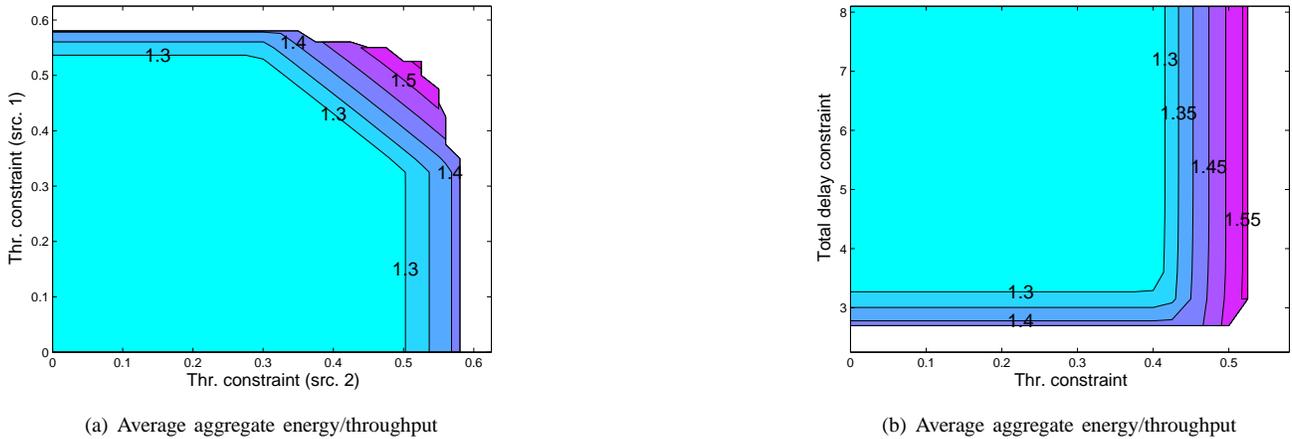

Fig. 4. average energy expense per unit of throughput.

Here, we consider the objective of minimizing energy expense normalized by throughput, under constraints on throughput, delay and packet loss. Numerical results obtained solving the linear fractional program presented in this work show how the system allocates transmissions as a function of the state of the network, enlightening interesting system behaviors.

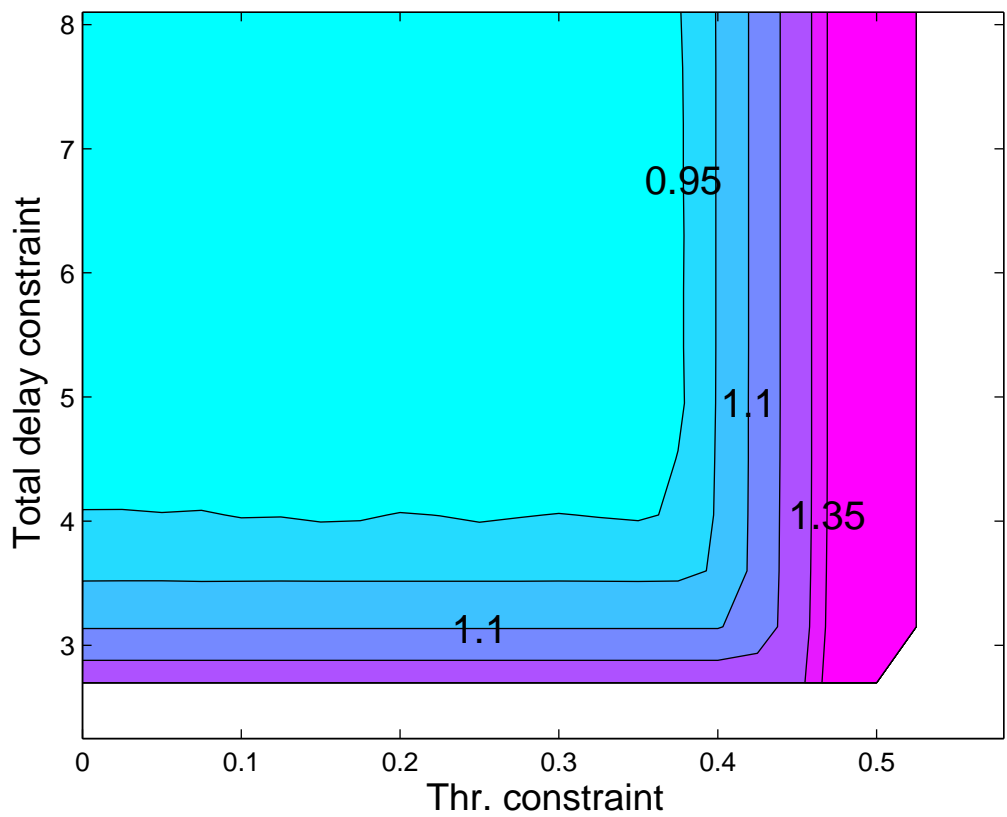

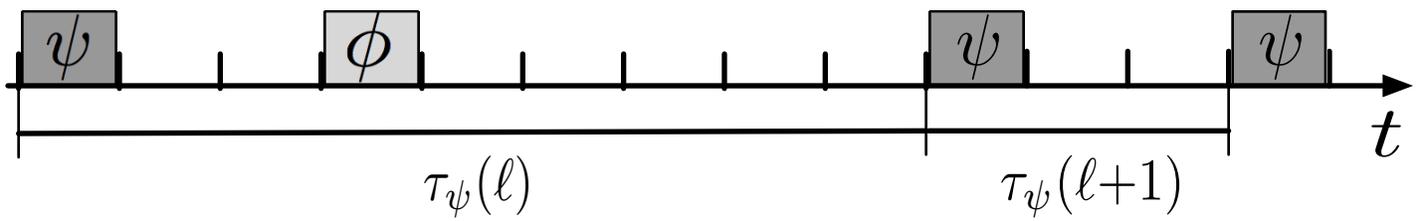

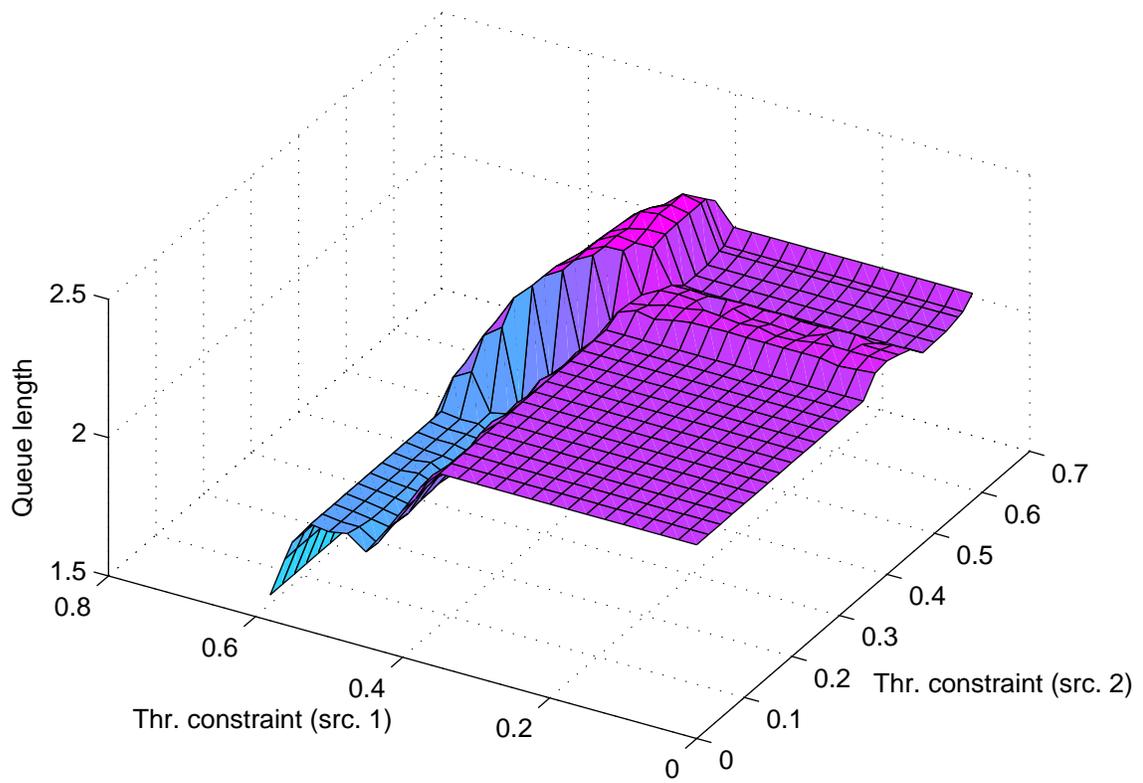

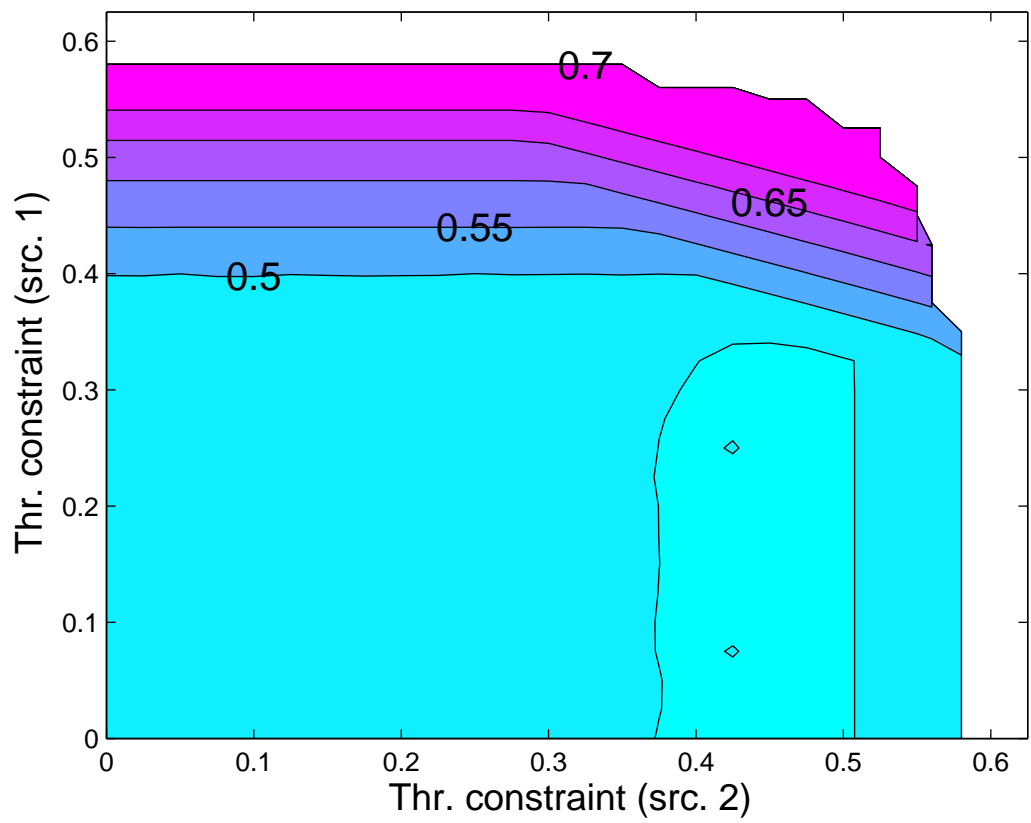

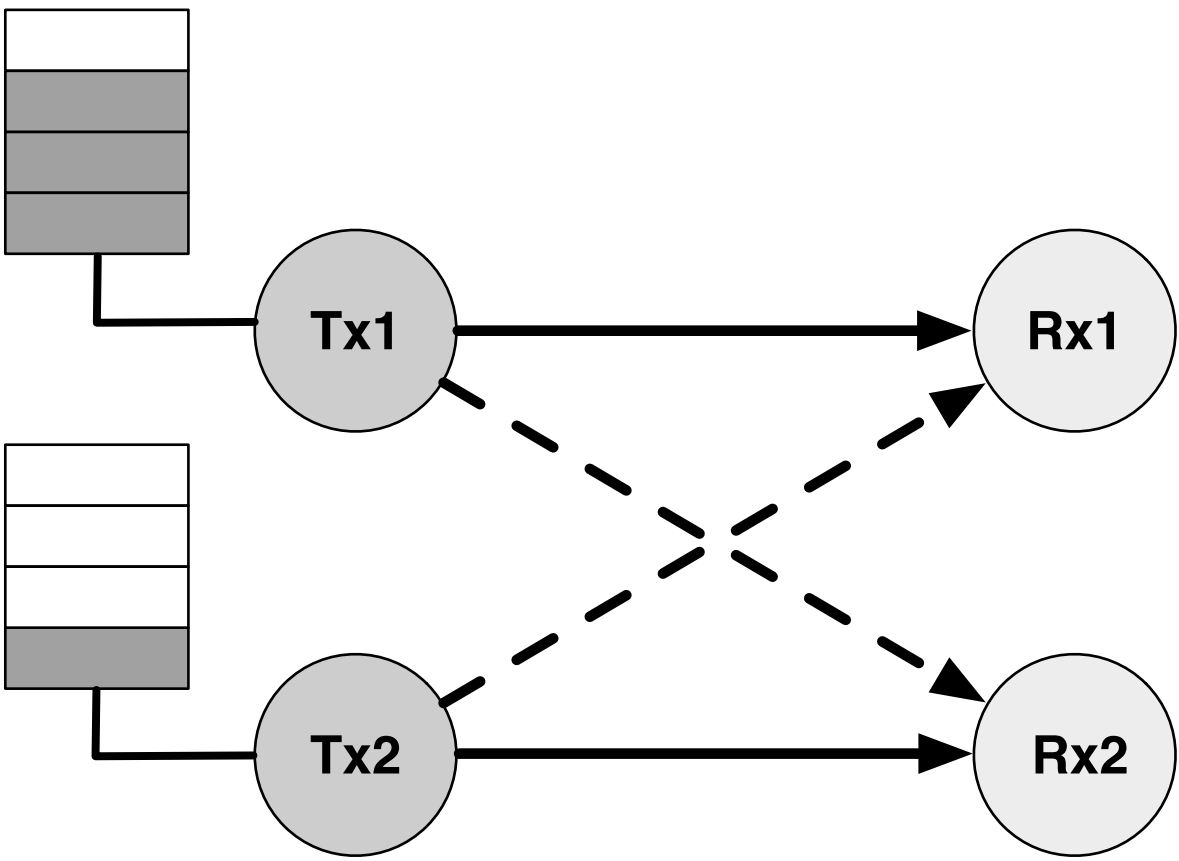